\documentclass[aps,pre,showpacs,amsmath,amssymb,reprint,dvips]{revtex4-1}

\usepackage{bm}
\usepackage[dvips]{graphicx}

\begin{document}

\title{
Integrated analysis of energy transfers in elastic-wave turbulence
}

\author{Naoto Yokoyama}
\email{yokoyama@kuaero.kyoto-u.ac.jp}
\affiliation{Department of Aeronautics and Astronautics, Kyoto University, Kyoto 615-8540, Japan}

\author{Masanori Takaoka}
\email{mtakaoka@mail.doshisha.ac.jp}
\affiliation{Department of Mechanical Engineering, Doshisha University, Kyotanabe 610-0394, Japan}

\date{\today}

\begin{abstract}
 In elastic-wave turbulence,
 strong turbulence appears in small wave numbers
 while weak turbulence does in large wave numbers.
 Energy transfers in the coexistence of these turbulent states
 are numerically investigated in both of the Fourier space and the real space.
 An analytical expression of a detailed energy balance reveals
 from which mode to which mode energy is transferred in the triad interaction.
 Stretching energy
 excited by external force
 is transferred nonlocally and intermittently
 to large wave numbers as the kinetic energy in the strong turbulence.
 In the weak turbulence,
 the resonant interactions according to the weak turbulence theory
 produces cascading net energy transfer to large wave numbers.
 Because the system's nonlinearity shows strong temporal intermittency,
 the energy transfers are investigated at active and moderate phases separately.
 The nonlocal interactions in the Fourier space are characterized by the intermittent bundles of fibrous structures in the real space.
\end{abstract}


\maketitle

\section{Introduction}

Investigation of energy transfers,
which reveals how the nonlinear interactions redistribute energy among scales,
is essential to understand turbulence dynamics.
For the Navier--Stokes turbulence,
the so-called K41 theory~\cite{K41} for energy cascade is based on 
the locality of the energy transfer.
Representing the Navier--Stokes equation in the Fourier space,
we can investigate the scale-by-scale energy budget.
The energy transfer is caused by a triad interaction in the Navier--Stokes turbulence.
When the three wave-number vectors, $\bm{k}$, $\bm{k}_1$ and $\bm{k}_2$, make
a triad, $\bm{k}+\bm{k}_1+\bm{k}_2=\bm{0}$,
in the incompressible fluid,
the sum of the energies of the three wave-number modes is conserved. 
The energy conservation through this triad interactions 
is called detailed energy balance.
The energy is transferred locally in the Fourier space,
because the small wave-number mode plays a role only as a mediator~\cite{:/content/aip/journal/pofa/2/3/10.1063/1.857736,*:/content/aip/journal/pofa/4/2/10.1063/1.858309,*:/content/aip/journal/pofa/2/9/10.1063/1.857818,:/content/aip/journal/pofa/4/4/10.1063/1.858296}.
The relation is under serious study
between the energy cascade in the Fourier space
and the dynamics of real-space structures.

It is important also in wave turbulence to identify
from which mode a wave-number mode in the inertial subrange obtains energy,
and to which mode the wave-number mode gives energy.
The energy transfer has not been fully understood yet,
because we are less successful even in obtaining the analytical expression for it
in such a way that is consistent with the energy conservation.
When the nonlinearity of wave turbulence is weak,
the weak turbulence theory well describes the transfer of the linear energy,
which is conserved by the kinetic equation.
Even then,
the detailed energy balance at each scale has never been investigated.
The difficulty comes in identifying the energy transfer between two wave-number modes
in the interactions among three, four or more wave-number modes.

Recently, in elastic-wave turbulence,
the analytical expression of the energy transfer has been reported in Ref.~\cite{PhysRevE.90.063004}. 
The elastic-wave turbulence has been used
for numerical or experimental verification
of the weak turbulence theory~\cite{during2006weak,mordant2008there,*boudaoud2008observation}.
In fact,
the weak turbulence theory is valid
in the large wave numbers where the nonlinearity is weak~\cite{during2006weak}.
The strong turbulence also appears in the small wave numbers,
and the weak turbulence and the strong turbulence coexist~\cite{PhysRevLett.110.105501,PhysRevE.89.012909}.
The analytical expression of the energy transfer in Ref.~\cite{PhysRevE.90.063004} requires no assumptions about the nonlinearity,
and hence it is exact and can be applied to the strong turbulence 
as well as the weak turbulence.

Such coexistence
reminds us of an important conjecture for the energy transfer,
i.e., critical balance in anisotropic turbulence~\cite{sridhar1994toward}.
The critical balance
also in isotropic turbulence such as gravity water waves and Kelvin waves
was suggested~\cite{nazarenkobook,*doi:10.1142/9789814366946_0006}.
The critical balance predicts that
the energy fluxes in the weak and strong turbulence are much different.
However,
the energy fluxes in the weak and strong turbulence in the elastic-wave turbulence
are not much different
because of the system's isotropy~\cite{PhysRevE.89.062925,PhysRevE.90.063004}.
The mechanism of the energy transfer in the coexistence of the weak and strong turbulence
in the isotropic systems
is needed to be understood carefully and quantitatively.

In the wave turbulence,
large-amplitude structures localized in space and time
are often found
when the nonlinearity is large.
An important example of the intermittent structures is the rogue wave,
which is formed in oceans, optics, superfluid Helium, plasmas and so on.
(See Ref.~\cite{Onorato201347} and references therein.)
The rogue wave is considered to be formed by the side-band instability
of the energy-containing modes~\cite{janssen_freak}.
The relation between such intermittent structures and energy transfer
has been claimed,
though direct evidence has not been found yet.
One of the difficulties in finding the evidence comes from the strong nonlinearity,
which requires the evaluation of the energy transfer
by including the nonlinear part of the energy in wave turbulence.

Such large-amplitude structures
are also of interest in the mechanics of elastic membrane.
When thin elastic sheets are deformed largely,
the focusing of excessive strain leads to almost singular structures
in the static crumpling of elastic membrane,
such as developable cones ({\itshape d}-cones) and ridges.
(See Ref.~\cite{RevModPhys.79.643} and references therein.)
The existence of dynamical crumpling,
named after the analogy with the above static crumpling,
are reported in Ref.~\cite{PhysRevLett.111.054302}
by simulating the unsteady elastic-wave turbulence.
Although the dynamical crumpling does not reach the singular structures,
the localized structures are remarkable for the strong nonlinearity.
These large-scale structures exist in the strong turbulence,
and they should give the mechanism of the energy transfer
different from the resonant interactions in the weak turbulence.
It is also observed in the elastic-wave turbulence
that the non-Gaussian statistics of the fluctuation exhibits small-scale intermittency~\cite{PhysRevE.94.011101}.

In this paper,
the energy transfers in the elastic-wave turbulence
are numerically investigated.
The energy transfers between kinetic energy of a wave-number mode
to stretching energy of another mode
are quantitatively evaluated
according to the triad interaction functions defined in Ref.~\cite{PhysRevE.90.063004}.
The energy transfers in the small wave numbers
are classified based on the system's nonlinearity,
and then the transfers due to the large-scale structures
are found to be much different from those due to the resonant interactions.
We also investigate the energy transfer in the real space
and the relation between the dynamics of the real-space structures
and the energy transfer in the Fourier space.
The transfers
reveal the different mechanism of the nonlinear interactions
in the weak and strong turbulence.

 \section{Governing equation and numerical simulations}

The governing equation of elastic waves propagating in a thin plate
is the F\"{o}ppl-von K\'{a}rm\'{a}n (FvK) equation.
The FvK equation is given as an equation for the displacement $\zeta$~\cite{llelasticity,audoly2010elasticity}
as follows.
 \begin{align}
  \rho\frac{\partial^2 \zeta}{\partial t^2}
  = - \frac{Yh^2}{12(1-\sigma^2)} \Delta^2 \zeta
  + \left\{ \zeta, \chi \right\}
  ,
  \label{eq:FvKzetaR}
 \end{align}
where the auxiliary variable $\chi$ is the Airy stress potential 
defined as
 \begin{align}
 \Delta^2 \chi = -\frac{Y}{2} \left\{\zeta, \zeta \right\}
 .
\label{eq:chiR}
 \end{align}
 Here,
 $\Delta$ represents the Laplace operator,
 and 
 \begin{align}
 \{f,g\} = 
 \frac{\partial^2 f}{\partial x^2} \frac{\partial^2 g}{\partial y^2}
 + \frac{\partial^2 f}{\partial y^2} \frac{\partial^2 g}{\partial x^2} 
 -2\frac{\partial^2 f}{\partial x \partial y} \frac{\partial^2 g}{\partial x \partial y}
 \end{align}
is the Monge--Amp\`ere operator.
The density $\rho$, the Young's modulus $Y$ and the Poisson ratio $\sigma$
are the material quantities of an elastic plate,
and $h$ denotes the thickness of the plate.

The momentum $p$ is defined as
 \begin{align}
  p=\rho\frac{\partial \zeta}{\partial t}.
  \label{eq:pR}
 \end{align}
 Then,
 under the periodic boundary condition,
the FvK equation (\ref{eq:FvKzetaR}), (\ref{eq:chiR}) and (\ref{eq:pR})
is rewritten as
\begin{subequations}
\begin{align}
&
\frac{d\zeta_{\bm{k}}}{dt} = \frac{p_{\bm{k}}}{\rho}
,
\quad
\frac{dp_{\bm{k}}}{dt}  = - \rho \omega_{\bm{k}}^2 \zeta_{\bm{k}}
 + \!\!\!\!
\sum_{\bm{k}_1+\bm{k}_2=\bm{k}}
\!\!\!\!
 |\bm{k}_1 \times \bm{k}_2|^2 \zeta_{\bm{k}_1} \chi_{\bm{k}_2}
,
\label{eq:zetapk}
 \\
&
\chi_{\bm{k}} = -\frac{Y}{2k^4} \sum_{\bm{k}_1+\bm{k}_2=\bm{k}} |\bm{k}_1 \times \bm{k}_2|^2 \zeta_{\bm{k}_1} \zeta_{\bm{k}_2}
,
\label{eq:chik}%
\end{align}%
\label{eq:FvKink}%
\end{subequations}%
where $\zeta_{\bm{k}}$, $p_{\bm{k}}$ and $\chi_{\bm{k}}$
are the Fourier coefficients of the displacement, of the momentum, and of the Airy stress potential, respectively.
The linear dispersion relation
gives the frequency $\omega_{\bm{k}}$
as
\begin{align}
 \omega_{\bm{k}} = \sqrt{\frac{Yh^2}{12(1-\sigma^2)\rho}} \ k^2
.
\label{eq:lineardispersion}
\end{align}

By introducing the complex amplitude
defined as
\begin{align}
 a_{\bm{k}} = 
\frac{\rho \omega_{\bm{k}} \zeta_{\bm{k}} + i p_{\bm{k}}}{\sqrt{2 \rho \omega_{\bm{k}}}}
,
\label{eq:ComplexAmplitude}
\end{align}
Eq.~(\ref{eq:FvKink}) is rewritten as
\begin{align}
 \frac{da_{\bm{k}}}{dt} 
=& - i \omega_{\bm{k}} a_{\bm{k}}
\nonumber\\
&
- \frac{iY}{8\rho^2}
\!\!
 \sum_{\bm{k}_1+\bm{k}_2+\bm{k}_3=\bm{k}}
\!\!\!\!\!\!\!\!
\frac{|\bm{k} \times \bm{k}_1|^2|\bm{k}_2 \times \bm{k}_3|^2}{|\bm{k}_2+\bm{k}_3|^4 }
\nonumber\\
& \quad \times
\frac{
(a_{\bm{k}_1} + a_{-\bm{k}_1}^{\ast})
(a_{\bm{k}_2} + a_{-\bm{k}_2}^{\ast})
(a_{\bm{k}_3} + a_{-\bm{k}_3}^{\ast})
}{\sqrt{\omega_{\bm{k}}\omega_{\bm{k}_1}\omega_{\bm{k}_2}\omega_{\bm{k}_3}}}
.
\label{eq:fvka}
\end{align}

Direct numerical simulation is performed
according to the following equation,
where the external force $F_{\bm{k}}$
and the dissipation $D_{\bm{k}}$
are added
to make statistically-steady non-equilibrium states:
\begin{align}
 \frac{da_{\bm{k}}}{dt} 
 = - i \omega_{\bm{k}} a_{\bm{k}}
 + \mathcal{N}_{\bm{k}}
 + F_{\bm{k}} + D_{\bm{k}}
 .
\label{eq:fvkaFD} 
\end{align}
Here, $\mathcal{N}_{\bm{k}}$
represents the second term in the right-hand side of Eq.~(\ref{eq:fvka}),
which shows the four-wave nonlinear interactions.
The external force $F_{\bm{k}}$ are artificially added
to the small wave numbers
so that $|a_{\bm{k}}|$ at $|\bm{k}| \leq 8\pi$
is constant in time.
The dissipation is added as $D_{\bm{k}} = -\nu |\bm{k}|^8 a_{\bm{k}}$,
which is effective in the wave-number range
$|\bm{k}| \gtrapprox 256\pi$
in the present work.

The direct numerical simulation according to Eq.~(\ref{eq:fvkaFD})
is performed
for a plate having the periodic boundary of $1$m$\times 1$m
according to Ref.~\cite{mordant2008there,*boudaoud2008observation},
and then
the two-dimensional wave-number vector $\bm{k}$ is discretized as $\bm{k} \in (2\pi \mathbb{Z})^2$.
The pseudo-spectral method,
where the number of the aliasing-free modes is $512 \times 512$,
is employed.
Details of the numerical scheme 
are explained in Ref.~\cite{PhysRevLett.110.105501}.

As a result of the numerical simulation,
the energy spectrum shows the coexistence of the weak turbulence and the strong turbulence~\cite{PhysRevLett.110.105501,PhysRevE.89.012909,PhysRevE.90.063004}.
The weakly nonlinear spectrum
which is a stationary solution of the kinetic equation~\cite{during2006weak},
is observed in the large wave numbers.
Another power law is
observed in the small wave numbers,
where the nonlinearity is relatively strong.
The separation wave number of the weak and strong turbulence
is located approximately at $k \approx 300$,
which will be found in the energy spectra and the energy transfers in Fig.~\ref{fig:phase} below.
We will investigate the energy transfers
in this coexistence.

 \section{Formulation of energy transfers}
\label{sec:formulation}
   \subsection{Energy transfers in Fourier space}
\label{ssec:formulationF}

In this subsection,
the energy transfers in the Fourier space
is reviewed mostly following to Ref.~\cite{PhysRevE.90.063004}.
The total energy of a wave-number mode $E_{\bm{k}}$ is give as
the sum of the kinetic energy $K_{\bm{k}}$,
the bending energy $V_{\mathrm{B} \bm{k}}$ and 
the stretching energy $V_{\mathrm{S} \bm{k}}$,
i.e., $E_{\bm{k}} = K_{\bm{k}} + V_{\mathrm{B} \bm{k}} + V_{\mathrm{S} \bm{k}}$.
Here,
\begin{align}
 K_{\bm{k}} = \frac{1}{2\rho} |p_{\bm{k}}|^2
 ,
 \quad
 V_{\mathrm{B} \bm{k}} = \frac{\rho \omega_{\bm{k}}^2}{2} |\zeta_{\bm{k}}|^2
 ,
 \quad
 V_{\mathrm{S} \bm{k}} = \frac{k^4}{2Y} |\chi_{\bm{k}}|^2
 .
\label{eq:defenergyies}%
\end{align}%
These energies give the Hamiltonian of the FvK equation~(\ref{eq:FvKink}):
\begin{align}
 \mathcal{H} =& \sum_{\bm{k}} \left(\frac{1}{2\rho} |p_{\bm{k}}|^2 + \frac{\rho \omega_{\bm{k}}^2}{2} |\zeta_{\bm{k}}|^2
 +\frac{k^4}{2Y} |\chi_{\bm{k}}|^2
  \right)
.
\label{eq:hamiltonian}
\end{align}
It must be noted that
the representation of the nonlinear part of the Hamiltonian
using the Airy stress potential $\chi_{\bm{k}}$
is not a mathematical trick.
Originally, the stretching energy is given by the strain, and hence the Airy potential,
which can be expressed by the displacement.
Thus, the notation by the Airy stress potential is more primitive
than that by the convolution of the displacement,
and is in accordance with the derivation.

The energy transfer for each of decomposed energy can be derived 
by taking the time-derivatives of Eq.~(\ref{eq:defenergyies}) 
and substituting Eq.~(\ref{eq:FvKink}) into them
(see also Ref.~\cite{PhysRevE.90.063004}):
\begin{subequations}
 \begin{align}
  T_{\mathrm{K} \bm{k}} &=
\frac{\hat{d} K_{\bm{k}}}{\hat{d}t} = T_{\mathrm{K} \bm{k}}^{(2)}+T_{\mathrm{K} \bm{k}}^{(4)}
  ,
  \\
T_{\mathrm{K} \bm{k}}^{(2)}
&=
 - \frac{\omega_{\bm{k}}^2}{2} p_{\bm{k}}^{\ast} \zeta_{\bm{k}}
 + \mathrm{c.c.}
 ,
 \label{eq:transferK2}
 \\
 T_{\mathrm{K} \bm{k}}^{(4)}
 &= 
 \frac{p_{\bm{k}}^{\ast}}{2\rho}
 \sum_{\bm{k}_1+\bm{k}_2=\bm{k}} \!\!\!\!
 |\bm{k}_1 \times \bm{k}_2|^2 \zeta_{\bm{k}_1} \chi_{\bm{k}_2}
 + \mathrm{c.c.}
 ,
 \label{eq:transferK4}
 \\
 T_{\mathrm{B} \bm{k}}
 &=
 \frac{\hat{d} V_{\mathrm{B} \bm{k}} }{\hat{d}t}
 =
 \frac{\omega_{\bm{k}}^2}{2} p_{\bm{k}}^{\ast} \zeta_{\bm{k}}
 + \mathrm{c.c.}
 ,
 \label{eq:transferVb}
 \\
T_{\mathrm{S} \bm{k}}
 &= 
 \frac{\hat{d} V_{\mathrm{S} \bm{k}} }{\hat{d}t}
 =
 - \frac{\chi_{\bm{k}}^{\ast}}{2\rho}
 \sum_{\bm{k}_1+\bm{k}_2=\bm{k}} \!\!\!\!
 |\bm{k}_1 \times \bm{k}_2|^2 p_{\bm{k}_1} \zeta_{\bm{k}_2} 
  + \mathrm{c.c.}
  ,
\label{eq:transferVs}%
 \end{align}%
\label{eq:energytransfers}%
\end{subequations}%
where $\hat{d}/\hat{d}t$ denotes the time-derivative in Eq.~(\ref{eq:FvKink}),
and the external force and the dissipation are disregarded here.
Note that Eq.~(\ref{eq:energytransfers}) is obtained directly
from Eq.~(\ref{eq:FvKink}) and Eq.~(\ref{eq:defenergyies})
without any approximations.
Then, the total-energy transfer is composed of these transfers as
$ T_{\bm{k}} = T_{\mathrm{K} \bm{k}} + T_{\mathrm{B} \bm{k}} + T_{\mathrm{S} \bm{k}}$.

The cancellation of
the quadratic part of the kinetic-energy transfer $T_{\mathrm{K} \bm{k}}^{(2)}$
given as Eq.~(\ref{eq:transferK2})
and the bending-energy transfer $T_{\mathrm{B} \bm{k}}$
given as Eq.~(\ref{eq:transferVb}),
$T_{\mathrm{K} \bm{k}}^{(2)}+T_{\mathrm{B} \bm{k}}=0$, corresponds to
exchange between the kinetic and potential energies of the identical wave number
as in the linear harmonic wave.
Note that the exchange between the kinetic and bending energies
in a wave number is referred to as transmutation shortly,
and the transmutation is distinguished
from the nonlinear energy transfer among different wave numbers.
The bending energy increases or decreases only through this linear transmutation.

The quartic part of the kinetic-energy transfer $T_{\mathrm{K} \bm{k}}^{(4)}$
 and the stretching-energy transfer $T_{\mathrm{S} \bm{k}}$
are the energy transfers
due to the nonlinear interactions among triads
$-\bm{k}+\bm{k}_1+\bm{k}_2=\bm{0}$.

By introducing the triad interaction functions,
\begin{subequations}
\begin{align}
 T_{\mathrm{K} \bm{k}\bm{k}_1\bm{k}_2}^{(4)} 
 &=
 \frac{|\bm{k}_1 \times \bm{k}_2|^2}{2\rho}
 p_{\bm{k}} \zeta_{\bm{k}_1} \chi_{\bm{k}_2} 
 \delta_{\bm{k}+\bm{k}_1+\bm{k}_2,\bm{0}}
 + \mathrm{c.c.}
 ,
 \\
 T_{\mathrm{S} \bm{k}\bm{k}_1\bm{k}_2}
 &=
 -\frac{|\bm{k}_1 \times \bm{k}_2|^2}{2\rho}
 \chi_{\bm{k}}  p_{\bm{k}_1} \zeta_{\bm{k}_2}
 \delta_{\bm{k}+\bm{k}_1+\bm{k}_2,\bm{0}}
 + \mathrm{c.c.}
 ,
\end{align}%
\label{eq:piecewisetriadfunc}%
\end{subequations}%
the nonlinear energy transfers can be written as
\begin{align}
T_{\mathrm{K} \bm{k}}^{(4)} 
=  \sum_{\bm{k}_1,\bm{k}_2} T_{\mathrm{K} \bm{k}\bm{k}_1\bm{k}_2}^{(4)} 
,
\quad
T_{\mathrm{S} \bm{k}} 
=\sum_{\bm{k}_1,\bm{k}_2} T_{\mathrm{S} \bm{k}\bm{k}_1\bm{k}_2}
.
\end{align}
It should be noted here that
the modes for $\bm{k}_1$ and $\bm{k}_2$
appearing in the triad interaction function of the total energy, $T_{\bm{k}\bm{k}_1\bm{k}_2}$, 
correspond to the different kinds of energies,
while those for the Navier--Stokes equation are identical.
For these triad interaction functions, 
a detailed energy balance can be derived
\begin{align}
&
 T_{\mathrm{K} \bm{k}\bm{k}_1\bm{k}_2}^{(4)} + T_{\mathrm{S} \bm{k}_2 \bm{k} \bm{k}_1}
 \nonumber\\
&
 =
 \frac{|\bm{k}_1 \times \bm{k}_2|^2}{2\rho}
 p_{\bm{k}} \zeta_{\bm{k}_1} \chi_{\bm{k}_2} 
 \delta_{\bm{k}+\bm{k}_1+\bm{k}_2,\bm{0}}
 + \mathrm{c.c.}
 \nonumber\\
 &
\qquad
 -\frac{|\bm{k} \times \bm{k}_1|^2}{2\rho}
 \chi_{\bm{k}_2} p_{\bm{k}} \zeta_{\bm{k}_1}
 \delta_{\bm{k}+\bm{k}_1+\bm{k}_2,\bm{0}}
 + \mathrm{c.c.}
 \nonumber\\
 &
 = 0
 .
 \label{eq:balanceKS}
\end{align}
Different from the transmutation of the quadratic part,
Eq.~(\ref{eq:balanceKS}) represents that
the stretching energy of $\bm{k}_2$ is transformed
into the kinetic energy of $\bm{k}$ and vice versa
through the triad interaction $\bm{k}+\bm{k}_1+\bm{k}_2=\bm{0}$.
It is of interest that
the mode $\zeta_{\bm{k}_1}$ in $T_{\mathrm{K} \bm{k}\bm{k}_1\bm{k}_2}^{(4)}$
and the mode $\zeta_{\bm{k}_1}$ in $T_{\mathrm{S} \bm{k}_2 \bm{k} \bm{k}_1}$
serves as a mediator
between these two kinds of energies in the nonlinear interactions.
This provides insight for gain or loss of the energies of the two wave-number modes
in the triad in the elastic-wave turbulence,
in marked contrast with the triad interaction in the Navier--Stokes turbulence.

The triad interaction function of the total energy,
$T_{\bm{k}\bm{k}_1\bm{k}_2}=T_{\mathrm{K} \bm{k}\bm{k}_1\bm{k}_2}^{(4)}+T_{\mathrm{S} \bm{k}\bm{k}_1\bm{k}_2}$,
satisfies the detailed energy balance
through the triad interaction $\bm{k}+\bm{k}_1+\bm{k}_2=\bm{0}$:
\begin{align}
T_{\bm{k}\bm{k}_1\bm{k}_2}+T_{\bm{k}_1\bm{k}_2\bm{k}}+T_{\bm{k}_2\bm{k}\bm{k}_1}=0.
\label{eq:detailedbalance}
\end{align}
Namely,
the triad interaction function shows the interchanges of the energy among the wave-number modes
keeping the sum of the energies of the three wave-number modes.

 \begin{figure}[t]
  \includegraphics[scale=.4]{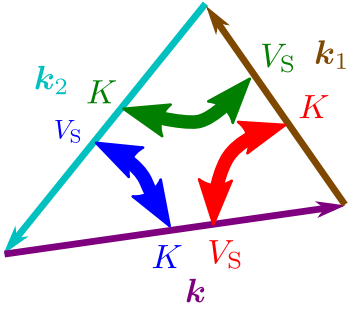}
  \caption{(Color online)
  Detailed energy balance due to triad interactions:
  $T_{\bm{k}\bm{k}_1\bm{k}_2}+T_{\bm{k}_1\bm{k}_2\bm{k}}+T_{\bm{k}_2\bm{k}\bm{k}_1}=0$.
  \label{fig:triad}
  }
 \end{figure}
 The two kinds of the detailed energy balance, Eqs.~(\ref{eq:balanceKS}) and (\ref{eq:detailedbalance}),
indicate that
the energy is transferred
by changing its form between the kinetic energy and the stretching energy
owing to the triad interaction.
The energy budget in the triad is schematically drawn in Fig.~\ref{fig:triad}.
It must be emphasized here again that 
though three wave-number modes appear in the triad interactions,
we can identify
from which mode or to which mode the energy is transferred
in the elastic-wave turbulence.
In other turbulent systems,
the energy transfer of a wave-number mode is obtained
only as the sum of the exchange with the multiple wave-number modes:
with two wave-number modes in a triad interaction
and with three modes in a quartet interactions.
Therefore, strong assumptions are usually required to identify the wave number
from or to which mode the energy is transferred.
For example, the locality of the nonlinear interactions was assumed in Ref.~\cite{WEBB1978279}.
Since the combination of $\zeta_{\bm{k}_1}$ and $\chi_{\bm{k}_2}$
and that of $p_{\bm{k}_1}$ and $\zeta_{\bm{k}_2}$
are used as elementary modes in the triad interaction functions~(\ref{eq:piecewisetriadfunc}),
and each mode corresponds to each wave number composing the triangle,
we can clearly distinguish between the roles of $\bm{k}_1$ and $\bm{k}_2$.
It is highly advantageous to the identification of the details of the energy transfer.

\subsection{Energy transfers in real space}
\label{ssec:formulationR}

In order to investigate the relation between the dynamics in the real space
and the energy transfers in the Fourier space,
we here define the energy transfers in the real space.
At each location in the real space,
the densities of the kinetic, bending and stretching energies
are given as
\begin{subequations}
 \begin{align}
  K(\bm{x}) &= \frac{1}{2\rho} p^2
  ,
  \\
  V_{\mathrm{B}}(\bm{x}) &=
  \frac{Yh^2}{24(1-\sigma^2)}
  \left[(\Delta\zeta)^2
  -(1-\sigma)\left\{\zeta,\zeta\right\}\right]
  ,
  \\
  V_{\mathrm{S}}(\bm{x}) &=
  \frac{1}{2Y}
  \left[(\Delta\chi)^2 -(1+\sigma)\left\{\chi,\chi\right\}\right]
.
\label{eq:VSreal}
 \end{align}
\end{subequations}
The bending energy derives from the out-of-plane displacement,
while the stretching energy comes from the in-plane strain.
See, for example, Ref.~\cite{audoly2010elasticity}.

Corresponding to Eq.~(\ref{eq:energytransfers}),
the decomposed energy transfers in the real space
are defined here as follows:
\begin{subequations}
 \begin{align}
\frac{\partial K}{\partial t}
 &=
 T_{\mathrm{K}}^{(2)}(\bm{x})
 + T_{\mathrm{K}}^{(4)}(\bm{x})
\nonumber\\
  &=
 - \frac{Yh^2}{12(1-\sigma^2)\rho} p\Delta^2\zeta
 +\frac{1}{\rho} p\left\{\zeta,\chi\right\}
  ,
  \label{eq:transferinRK}%
  \\
 \frac{\partial V_{\mathrm{B}}}{\partial t}
 &=
 T_{{\mathrm{B}}}^{(2)}(\bm{x})
 + T_{{\mathrm{B}}}^{(\mathrm{D})}(\bm{x})
  \nonumber\\
  &=
\frac{Yh^2}{12(1-\sigma^2)\rho} (\Delta\zeta) (\Delta p)
 - \frac{Yh^2}{12(1+\sigma)\rho} \left\{\zeta,p\right\}
,
\label{eq:transferinRB}%
  \\
 \frac{\partial V_{\mathrm{S}}}{\partial t}
 &=
 T_{{\mathrm{S}}}^{(4)}(\bm{x})
 + T_{{\mathrm{S}}}^{(\mathrm{D})}(\bm{x})
\nonumber\\
  &=
 - \frac{1}{\rho} (\Delta\chi)(\Delta^{-1}\left\{\zeta,p\right\})
 + \frac{1+\sigma}{\rho}
\left\{\chi,\Delta^{-2}\left\{\zeta,p\right\}\right\}
.
\label{eq:transferinRS}%
 \end{align}%
\label{eq:transferinR}%
\end{subequations}%
Note that both $T_{{\mathrm{B}}}^{(\mathrm{D})}(\bm{x})$ and $T_{{\mathrm{S}}}^{(\mathrm{D})}(\bm{x})$
have divergence forms,
because the Monge--Amp\`ere operator can be rewritten as
\begin{align}
  \{f,g\} =& 
  \frac{\partial}{\partial x}\left(
    \frac{\partial f}{\partial x} \frac{\partial^2 g}{\partial y^2}
  - \frac{\partial f}{\partial y} \frac{\partial^2 g}{\partial x \partial y}
  \right)
\nonumber\\
  &+ \frac{\partial}{\partial y}\left(
     \frac{\partial f}{\partial y} \frac{\partial^2 g}{\partial x^2} 
   - \frac{\partial f}{\partial x} \frac{\partial^2 g}{\partial x \partial y} 
  \right) .
\end{align}

The representation that is more similar to Eq.~(\ref{eq:energytransfers})
can be obtained by integrating these expressions over a finite area $A$
and by using partial integral:
\begin{align}
 \frac{\partial {\cal{K}}}{\partial t}
 &= T_{{\cal{K}}}^{(2)}     + \ T_{{\cal{K}}}^{(4)}
\\
 &= -\frac{Yh^2}{12(1-\sigma^2)\rho}
 \int_{A} p\Delta^2\zeta dA
 +\frac{1}{\rho}\int_{A} p\left\{\zeta,\chi\right\} dA
,
 \nonumber\\
 \frac{\partial {\cal{V}}_{\mathrm{B}}}{\partial t}
 &=T_{{\mathcal{B}}}^{(2)}
 + T_{{\mathcal{B}}}^{(\mathrm{D})}
 \nonumber\\
 &=
 \frac{Yh^2}{12(1-\sigma^2)\rho}
 \int_{A} p\Delta^2\zeta dA  + {\text{B.V.}}(\partial A)
,
 \nonumber\\
 \frac{\partial {\cal{V}}_{\mathrm{S}}}{\partial t}
 &=T_{{\mathcal{S}}}^{(4)}
 + T_{{\mathcal{S}}}^{(\mathrm{D})}
 \nonumber\\
 &=  -\frac{1}{\rho}\int_{A} \chi \left\{ p, \zeta\right\} dA 
 + {\text{B.V.}}(\partial A)
,
\end{align}%
where B.V.$(\partial A)$ stands for the boundary values
surrounding the finite area $A$.
These expressions more clearly show the energy balance
among the kinetic, bending, and stretching energies
than those in Eq.~(\ref{eq:transferinR}).
For example, a similar expression corresponding to 
the detailed energy balance, Eq.~(\ref{eq:balanceKS}), can be written as
\begin{align}
  T_{{\cal{K}}}^{(4)} + T_{{\mathcal{S}}}^{(4)}
=& \frac{1}{\rho}\int_{A} p\left\{\zeta,\chi\right\} dA
-\frac{1}{\rho}\int_{A} \chi \left\{ p, \zeta\right\} dA 
\nonumber\\
=& {\text{B.V.}}(\partial A) 
.
\end{align}
When the area $A$ is the whole domain,
we obtain ${\text{B.V.}}(\partial A)=0$ 
owing to the periodic boundary condition.

\section{Results}
\label{sec:results}

\subsection{Detailed energy transfers among wave-number modes}

\begin{figure}[t]
 \includegraphics[scale=.9]{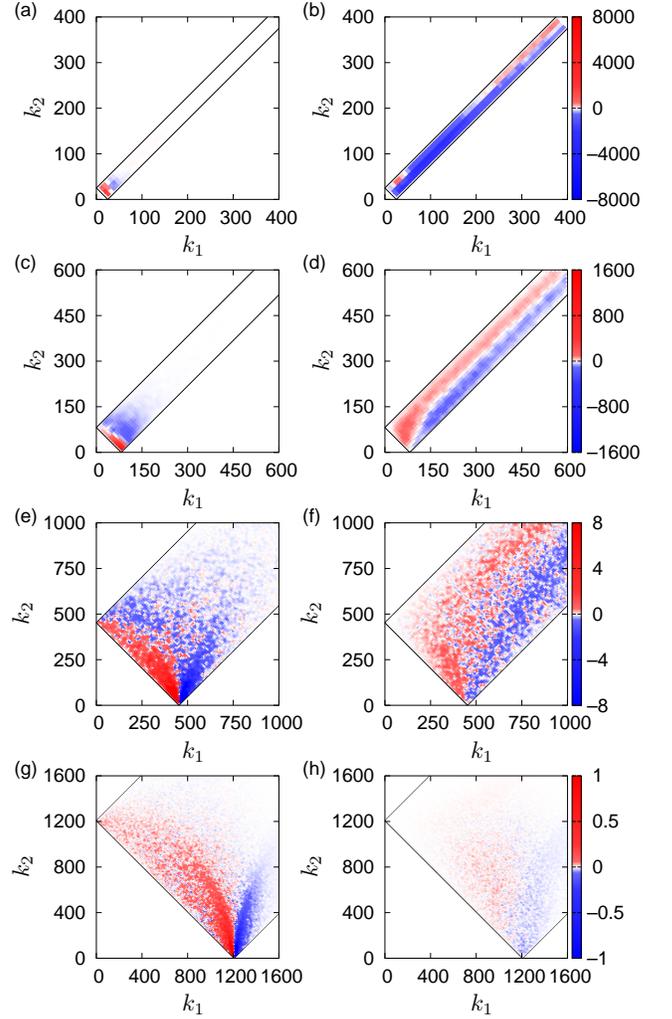}
 \caption{
 (Color)
 Energy transfer functions.
 (a) and (b): to $|\bm{k}|=8\pi$,
 (c) and (d): to $|\bm{k}|=26\pi$,
 (e) and (f): to $|\bm{k}|=144\pi$,
 and
 (g) and (h): to $|\bm{k}|=384\pi$.
 (a), (c), (e) and (g): to kinetic energy $\mathcal{T}_{\mathrm{K} \bm{k}}^{(4)}(k_1,k_2)$,
 where $k_1$ and $k_2$ are respectively the wave numbers for $\zeta$ and $\chi$
 as defined in Eq.~(\ref{eq:energytransferfunctionK}),
 and
 (b), (d), (f) and (h): to stretching energy $\mathcal{T}_{\mathrm{S} \bm{k}}(k_1,k_2)$,
 where $k_1$ and $k_2$ are respectively the wave numbers for $p$ and $\zeta$
 as defined in Eq.~(\ref{eq:energytransferfunctionS}).
 \label{fig:dbave}
 }
\end{figure}

To investigate the detailed energy balance through the triad interaction functions (see Eq.~(\ref{eq:balanceKS})),
the azimuthally-integrated energy transfer functions are defined as
\begin{subequations}
  \begin{align}
   &
  \mathcal{T}_{\mathrm{K} \bm{k}}^{(4)}(k_1,k_2)
   =
   \frac{1}{\Delta k_1 \Delta k_2}
 {\sum_{\bm{k}_1^{\prime},\bm{k}_2^{\prime}}}^{\prime}
 T_{\mathrm{K} \bm{k}\bm{k}_1^{\prime}\bm{k}_2^{\prime}}^{(4)} 
  \nonumber\\
  &=
   \frac{1}{\Delta k_1 \Delta k_2}
   {\sum_{\bm{k}_1^{\prime},\bm{k}_2^{\prime}}}^{\prime}
 \frac{|\bm{k}_1^{\prime} \times \bm{k}_2^{\prime}|^2}{2\rho}
 p_{\bm{k}} \zeta_{\bm{k}_1^{\prime}} \chi_{\bm{k}_2^{\prime}} 
 \delta_{\bm{k}+\bm{k}_1^{\prime}+\bm{k}_2^{\prime},\bm{0}}
 + \mathrm{c.c.}
   ,
   \label{eq:energytransferfunctionK}
   \\
   &
  \mathcal{T}_{\mathrm{S} \bm{k}}(k_1,k_2)
   =
   \frac{1}{\Delta k_1 \Delta k_2}
  {\sum_{\bm{k}_1^{\prime},\bm{k}_2^{\prime}}}^{\prime}
 T_{\mathrm{S} \bm{k}\bm{k}_1^{\prime}\bm{k}_2^{\prime}}
  \nonumber\\
  &=
   -
   \frac{1}{\Delta k_1 \Delta k_2}
  {\sum_{\bm{k}_1^{\prime},\bm{k}_2^{\prime}}}^{\prime}
 \frac{|\bm{k}_1^{\prime} \times \bm{k}_2^{\prime}|^2}{2\rho}
 \chi_{\bm{k}}  p_{\bm{k}_1^{\prime}} \zeta_{\bm{k}_2^{\prime}}
 \delta_{\bm{k}+\bm{k}_1^{\prime}+\bm{k}_2^{\prime},\bm{0}}
 + \mathrm{c.c.}
 ,
   \label{eq:energytransferfunctionS}
  \end{align}%
\end{subequations}%
where
${\sum_{\bm{k}_1^{\prime},\bm{k}_2^{\prime}}}^{\prime}$
denotes
the summation over
$|k_1^{\prime} - k_1| < \Delta k_1/2$
and
$|k_2^{\prime} - k_2| < \Delta k_2/2$.
Here, $\Delta k_1=\Delta k_2=2\pi$ is employed.
The energy transfer functions shown in the following figures~\ref{fig:dbave} and \ref{fig:intave}
are obtained by averaging over 
$1024$ realizations, $4$ different times, and directions of $\bm{k}$.
The time interval of the $4$ different times is sufficiently longer than the longest linear period 
to validate the statistical independence.
The numbers of the samples averaged over the directions $\bm{k}$ are at least $6$,
which is for $k=8\pi$.
We compared the result
with those obtained by reduced numbers of the samples and by different $\Delta k_1$ and $\Delta k_2$,
and the result is confirmed to be robust.

These energy transfer functions
are drawn in Fig.~\ref{fig:dbave} for four representative wave numbers.
The wave number $|\bm{k}|=8\pi \approx 25$ is in the forcing range.
Both of the wave numbers $26\pi \approx 82$ and $144\pi \approx 450$ are in the inertial subrange;
the former is in the strong turbulence range,
and the latter in the weak turbulence range.
The wave number $384\pi \approx 1200$ is in the dissipation range.
Note that the area is restricted within a diagonal rectangle
corresponding to the triangle inequality,
i.e., $|k_1-k_2|< k < k_1+k_2$.
The magnitude of $|\bm{k}|$ corresponds to the corner of the diagonal rectangle.

For the wave numbers $|\bm{k}|=8\pi$ (Fig.~\ref{fig:dbave}(a)),
the kinetic-energy transfer is positive for the wave numbers smaller than itself.
Namely,
$K_{\bm{k}}$ obtains the energy from $V_{\mathrm{S} \bm{k}_2}$
where $k_2 < k$.
The stretching energy is distributed over broad wave numbers
whose magnitude are $50 \lessapprox k_1 \lessapprox 200$.
(Fig.~\ref{fig:dbave}(b))
It exhibits that
the energy is transferred comparatively nonlocally to the wave numbers much larger than $8\pi$.
The sign of the stretching-energy transfer is always negative 
in $50 \lessapprox k_1 \lessapprox 200$.
Thus, the nonlocal transfers are statistically significant.

As shown in Fig.~\ref{fig:dbave}(c),
the wave numbers in the inertial subrange of the strong turbulence,
$|\bm{k}|=26\pi$,
obtain the energy as its kinetic energy nonlocally from small $k_2$.
The energy transfer from $k_2 \approx 0$ is substantial
because the sign of the energy transfer is always positive near the corner.
It is consistent with the loss of the stretching energy of the wave numbers in the forcing range
shown in Fig.~\ref{fig:dbave}(b).
On the other hand,
$V_{\mathrm{S} \bm{k}}$ obtains the energy
from $k_1$ when $k_1 < k$ or $k_1 < k_2$,
while it gives the energy
to $V_{\mathrm{S} \bm{k}_1}$ when $k_1 > k, k_2$
(Fig.~\ref{fig:dbave}(d)).

The wave number in the inertial subrange of the weak turbulence,
$|\bm{k}|=144\pi$,
has both positive and negative values of the kinetic-energy transfer
near the bottom corner where $k_2 \ll k$ (Fig.~\ref{fig:dbave}(e)).
It is in contrast with the positive values 
near the corner for the wave numbers in the forcing range and strong turbulence (Figs.~\ref{fig:dbave}(a) and (c)).
The energy transfers are consistent with
the negative and positive values at $k_1 = 144\pi \approx 450$ in Figs.~\ref{fig:dbave}(b) and (d).
The energy transfer function of the stretching energy
at $|\bm{k}|=144\pi$ (Fig.~\ref{fig:dbave}(f))
is similar with that at $|\bm{k}|=26\pi$ (Fig.~\ref{fig:dbave}(d))
because the self-similar cascading transfer is dominant.
This self-similarity makes similar structures
in the energy transfer functions at $|\bm{k}|=384\pi$ in the dissipation range.
(Figs.~\ref{fig:dbave}(g) and (h))

According to Eqs.~(\ref{eq:piecewisetriadfunc}) and (\ref{eq:balanceKS}),
the energy is transferred
between the kinetic energy of a wave-number mode and the stretching energy of another wave-number mode
owing to the nonlinear interactions.
The wave number of $\zeta$ which consists of a triad
mediates the energy transfers between the kinetic energy and stretching energy
without changing its bending energy by the nonlinear interactions.
Thus, the wave number of $\zeta$ does not contribute directly
to the redistribution of energy among the modes.
By virtue of this remarkable nature of the nonlinear interactions in the elastic-wave turbulence,
integration over the wave numbers of $\zeta$
clarifies the energy transfer
between the kinetic energy of a wave number and the stretching energy of another wave number.
We here refer to the energy transfer functions
integrated over the wave numbers of $\zeta$
as binary energy transfers.
The binary energy transfer $\mathcal{T}_{\mathrm{K}\bm{k}}^{(4)}$
to the kinetic energy of $\bm{k}$ from the stretching energy of $k_2$
and $\mathcal{T}_{\mathrm{S}\bm{k}}$ 
to the stretching energy of $\bm{k}$ from the kinetic energy of $k_1$
are, respectively, defined as
\begin{subequations}
\begin{align}
\mathcal{T}_{\mathrm{K}\bm{k}}^{(4)}(k_2)
 &= \int \mathcal{T}_{\mathrm{K} \bm{k}}^{(4)}(k_1,k_2) dk_1
 =
 \frac{1}{\Delta k_2}
 {\sum_{\bm{k}_2^{\prime}}}^{\prime} \sum_{\bm{k}_1^{\prime}}
  T_{\mathrm{K} \bm{k}\bm{k}_1^{\prime}\bm{k}_2^{\prime}}^{(4)} 
 ,
 \\
 \mathcal{T}_{\mathrm{S}\bm{k}}(k_1)
 &= \int \mathcal{T}_{\mathrm{S} \bm{k}}(k_1,k_2) dk_2
 =
 \frac{1}{\Delta k_1}
 {\sum_{\bm{k}_1^{\prime}}}^{\prime} \sum_{\bm{k}_2^{\prime}}
  T_{\mathrm{S} \bm{k}\bm{k}_1^{\prime}\bm{k}_2^{\prime}}
,
\end{align}
\end{subequations}
where
${\sum_{\bm{k}_1^{\prime}}}^{\prime}$
and
${\sum_{\bm{k}_2^{\prime}}}^{\prime}$
respectively denote
the summations over
$|k_1^{\prime} - k_1| < \Delta k_1/2$
and
$|k_2^{\prime} - k_2| < \Delta k_2/2$.

\begin{figure}[t]
 \includegraphics[scale=.7]{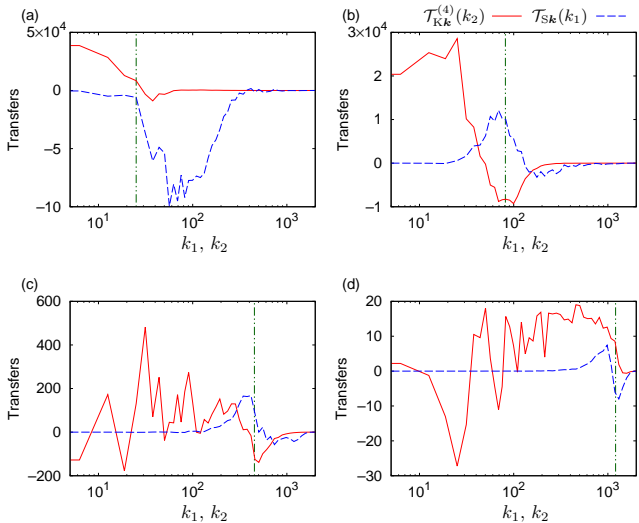}
 \caption{(Color online)
 Binary energy transfer functions $\mathcal{T}_{\mathrm{K}\bm{k}}^{(4)}(k_2)$ and $\mathcal{T}_{\mathrm{S}\bm{k}}(k_1)$
 from wave numbers $k_1$ or $k_2$
  (a) to $|\bm{k}|=8\pi$,
 (b) to $|\bm{k}|=26\pi$,
 (c) to $|\bm{k}|=144\pi$
 and
 (d) to $|\bm{k}|=384\pi$.
 The vertical lines show the representative wave numbers $|\bm{k}|$.
 The abscissa is scaled logarithmically.
 \label{fig:intave}
 }
\end{figure}

The binary energy transfers to the four representative wave numbers
are drawn in Fig.~\ref{fig:intave},
where $\Delta k_i$ is $2\pi$ for $k_i<26\pi$
and logarithmically-scaled for $k_i\ge 26\pi$.
In forcing range, $|\bm{k}|=8\pi$ (Fig.~\ref{fig:intave}(a)),
the mode obtains the energy as the kinetic energy
from the smaller wave numbers
as known from the fact that
the transfer of the kinetic energy ${\mathcal{T}}_{\mathrm{K} \bm{k}}^{(4)}(k_1,k_2)$
is large for small $k_2$. (Fig.~\ref{fig:dbave}(a)) 
Since the transfer of the stretching energy, $\mathcal{T}_{\mathrm{S}\bm{k}}(k_1)$,
is negatively large in
the range $50 \lessapprox k_1 \lessapprox 200$,
$V_{\mathrm{S} \bm{k}}$ at this scale is transferred
nonlocally to $K_{\bm{k}_1}$ in this range.
The nonlocal interactions are not fluctuations but statistically significant
as seen in Fig.~\ref{fig:dbave}(b).

In the inertial subrange of the strong turbulence,
$|\bm{k}|=26\pi$ (Fig.~\ref{fig:intave}(b)),
the wave number obtains the kinetic energy nonlocally 
from the wave numbers in the forcing range $k_2\leq 8\pi$.
It is consistent with the loss of $V_{\mathrm{S} \bm{k}}$
in the forcing range as shown in Fig.~\ref{fig:intave}(a).
The fact that the energy transfers from the forcing range
are comparable in Figs.~\ref{fig:intave}(a) and (b)
supports the large redistribution of the energy over broad wave numbers.
The transfer from $K_{\bm{k}}$ 
to $V_{\mathrm{S} \bm{k}_2}$ at $k_2$ slightly larger than $26\pi$,
is also noticeable.
The stretching energy at this scale $V_{\mathrm{S} \bm{k}}$
obtains the energy from $K_{\bm{k}_1}$ where $k_1$ is slightly smaller than $26\pi$,
and gives the energy to $K_{\bm{k}_1}$ where $150 \lessapprox k_1 \lessapprox 300$.
These interactions between comparable-scale modes
repeatedly transfer the kinetic and stretching energies
from the smaller to larger wave numbers,
and locally cascades energy step by step.
Therefore,
the nonlocal interactions with the wave-number modes in the forcing range
and the local interactions between comparable-scale modes
coexist in the strong turbulence.
The stretching-energy transfer, $\mathcal{T}_{\mathrm{S}\bm{k}}(k_1)$, 
for $k_1 \gtrapprox 400$ is almost $0$.
It indicates that
the positive and negative stretching-energy transfers
in the large wave numbers $k_1, k_2 \gg k$ in Fig.~\ref{fig:dbave}(d)
cancel each other.
The cancellation indicates that such nonlocal interactions do not cause net energy transfer.
It represents the ``drift'' of the small-scale wave $\bm{k}_1$
due to the large-scale wave of $\bm{k}$
resulting mainly in the small change of the direction of the small-scale wave.
It is similar with the sweeping effect known in the homogeneous isotropic Navier--Stokes turbulence~\cite{:/content/aip/journal/pofa/4/4/10.1063/1.858296}.
It must be noted again that
the nonlocal interactions between $K_{\bm{k}}$ where $\bm{k}$ is in the strong turbulence range
and $V_{\mathrm{S} \bm{k}}$ where $\bm{k}$ is in the forcing range
cause the net energy transfer,
and such statistically-significant nonlocal interactions are different from the sweeping effect.

The wave number in the inertial subrange of the weak turbulence
$|\bm{k}|=144\pi$ (Fig.~\ref{fig:intave}(c))
and that in the dissipation range $|\bm{k}|=384\pi$ (Fig.~\ref{fig:intave}(d))
show the locally-cascading kinetic- and stretching-energy transfers
in the same manner as the local interactions in the strong turbulence.
The nonlocal interactions with the forced wave numbers
for the kinetic-energy transfers
much fluctuate, and are not statistically significant.
The fluctuations are caused by the large positive and negative values
near the corner $k_2 \approx 0$ in Figs.~\ref{fig:dbave}(e) and (g),
and should be statistically canceled.
The tendency of the cancellation was confirmed
by comparison with the fluctuation in the fewer samples.
It represents the drift of the small-scale wave $\bm{k}$
due to the large-scale wave $\bm{k}_2$,
and it is the counterpart of the drift observed in large wave numbers in Fig.~\ref{fig:intave}(b).
Therefore,
though we observe that the large energy transfers from the wave numbers in
$k_1^2 + k_2^2 < k^2$ and $k_1+k_2 > k$
in Figs.~\ref{fig:dbave}(e) and (g),
only these local interactions between comparable-scale modes in the wave-number range
contribute the net energy transfers.
It should be noted here that 
the scale of the vertical axes of Figs.~\ref{fig:intave}(c) and (d) is respectively
smaller by two and three orders of magnitude
than those of Figs.~\ref{fig:intave}(a) and (b).
Even if there exist similar fluctuations in Figs.~\ref{fig:intave}(a) and (b),
they cannot be observed there owing to the substantial energy transfers.

 \begin{figure}[t]
  \includegraphics[scale=.6]{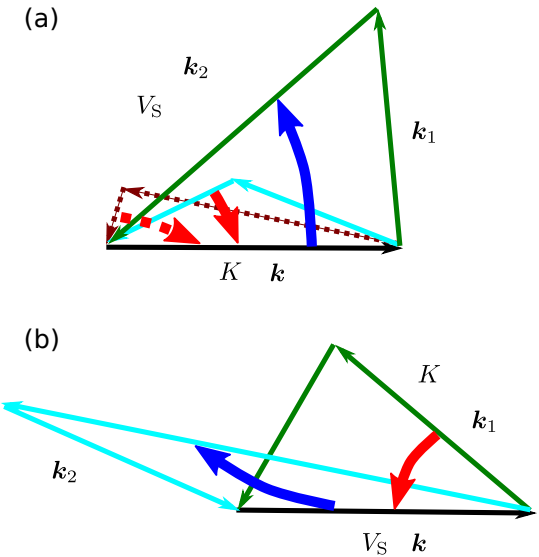}
  \caption{(Color online)
  Schematic net energy transfers due to the local interactions (solid arrows)
  and that due to the nonlocal interactions (dotted arrows).
  (a) $T_{\mathrm{K} \bm{k}\bm{k}_1\bm{k}_2}^{(4)}$
  and
  (b) $T_{\mathrm{S} \bm{k}\bm{k}_1\bm{k}_2}$.
  \label{fig:triaddb}
  }
 \end{figure}
 From Figs.~\ref{fig:dbave} and \ref{fig:intave},
 we can extract the shape of the triad interaction for the net energy transfers,
 which are schematically summarized in Fig.~\ref{fig:triaddb}.
 The thick arrows represent
 the energies from or to the wave number $\bm{k}$.
 Through the triad interactions of $T_{\bm{k}\bm{k}_1\bm{k}_2}+T_{\bm{k}_1\bm{k}_2\bm{k}}+T_{\bm{k}_2\bm{k}\bm{k}_1}=0$,
 $K_{\bm{k}}$ whose wave number $\bm{k}$ is in the inertial subrange
 is transformed from or into $V_{\mathrm{S}\bm{k}_2}$ (Fig.~\ref{fig:triaddb}(a)),
 while
 $V_{\mathrm{S}\bm{k}}$
 is transformed from or into $K_{\bm{k}_1}$ (Fig.~\ref{fig:triaddb}(b)).
 Note that only representative triads are shown in this figure.
 In particular,
 the lengths of the wave numbers of $\zeta$,
$\bm{k}_1$ in Fig.~\ref{fig:triaddb}(a) and $\bm{k}_2$ in Fig.~\ref{fig:triaddb}(b),
 are comparatively variable,
 which indicates that the triangle can be obtuse-angled as well as acute-angled.
 
 In Fig.~\ref{fig:triaddb}(a),
 the wave number $\bm{k}$
 that is either in the weak turbulence and the strong turbulence 
 obtains the kinetic energy from the stretching energy of $\bm{k}_2$
 through the triad that is
 $k_1, k_2 \approx k/2$ and $k_1$ slightly larger than $k_2$.
 (flattened triangle)
 On the other hand,
 $K_{\bm{k}}$ gives the energy to $V_{\mathrm{S} \bm{k}_2}$
 through the triad where $k_2$ is slightly larger than $k$.
 (nearly-equilateral triangle)
 As written above, the length of $\bm{k}_1$ is comparatively variable.
 These two transfers are caused by the local interactions in the Fourier space,
 and observed all over the wave numbers,
 The thick dotted arrow represents
 the nonlocal kinetic-energy transfer directly
 from the stretching energy of the small wave number $\bm{k}_2$
 that is in the forcing range.
 The nonlocal transfer emerges
 only if the wave number $\bm{k}$ is in the strong turbulence.
 (dotted triangle)

 The stretching-energy transfer $T_{\mathrm{S} \bm{k}\bm{k}_1\bm{k}_2}$ shown in Fig.~\ref{fig:triaddb}(b)
 has only the local interactions.
 Corresponding to the kinetic-energy transfer,
 $V_{\mathrm{S} \bm{k}}$ decreases
 by giving the energy to $K_{\bm{k}_1}$ where $k_1 \approx 2k$.
 (flattened triangle)
 In the same manner,
 $V_{\mathrm{S} \bm{k}}$ increases
 by obtaining the energy as $K_{\bm{k}_1}$
 whose $k_1$ slightly larger than $k$.
 (nearly-equilateral triangle)
 Similarly to the kinetic-energy transfer $T_{\mathrm{K} \bm{k}\bm{k}_1\bm{k}_2}$,
 the locally-cascading energy transfers are observed
 all over the wave numbers.

\subsection{Energy transfers in active and moderate phases}

\begin{figure}[t]
 \includegraphics[scale=.8]{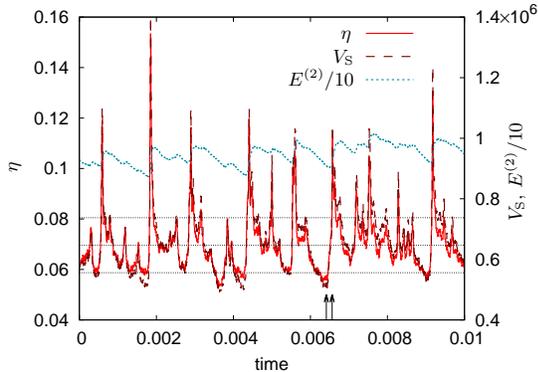}
 \caption{(Color online)
 Times series of nonlinearity $\eta$ (left axis),
 and linear energy $E^{(2)}$ and nonlinear energy $V_{\mathrm{S}}$.
 The linear energy is divided by $10$ for visibility.
 The mean of $\eta$ and $\pm 1$-sigma levels
 are represented by dotted lines.
 Two arrows at bottom indicate the representative times:
 moderate phase and active phase.
 \label{fig:tsnon}
 }
\end{figure}

The typical time series of the nonlinearity
as well as the linear and nonlinear energies
are drawn in Fig.~\ref{fig:tsnon}.
The intermittency leads us to investigate the energy transfer
by dividing the phases of the system into active and moderate phases.
The active and moderate phases
are classified by the nonlinearity of the system.

The nonlinearity is defined
as the ratio of the nonlinear energy to the linear energy:
$\eta = V_{\mathrm{S}}/E^{(2)} = \sum_{\bm{k}} V_{\mathrm{S}\bm{k}} / \sum_{\bm{k}}(K_{\bm{k}} + V_{\mathrm{B}\bm{k}})$.
The nonlinear energy (dashed curve) shows strong temporal intermittency,
which has sawtooth-wave profile with sudden jumps and slow relaxations, 
while the jumps of the linear energy (dotted curve) are synchronized,
but are not so large.
It results in resemblance between
strong temporal intermittency of the nonlinear energy 
and the that of the nonlinearity (solid curve).
In this work,
the active phases
when the system has strong nonlinearity
are defined so that
$\eta > \langle \eta \rangle + \sqrt{\langle (\eta - \langle \eta \rangle)^2 \rangle}$,
while the moderate phases are the phases when
$\eta < \langle \eta \rangle - \sqrt{\langle (\eta - \langle \eta \rangle)^2 \rangle}$.
Here, $\langle\cdot\rangle$ represents averaging over ensembles,
and $4096$ fields are used as ensembles.
As a result,
the numbers of statistically-independent wave fields in the active phase and in the moderate phase
are respectively $521$ and $393$.

\begin{figure}[t]
 \includegraphics[scale=.8]{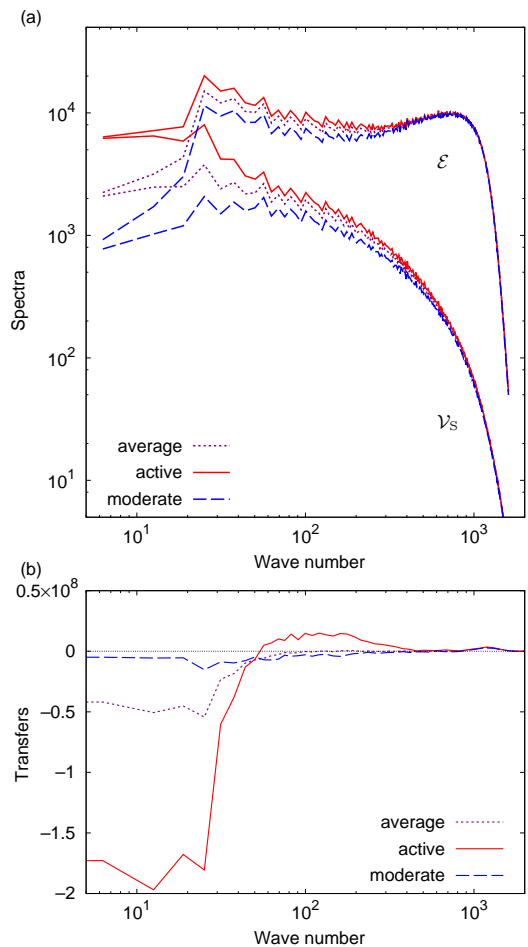}
 \caption{(Color online)
 (a): Total-energy spectra $\mathcal{E}$ and stretching-energy spectra $\mathcal{V}_{\mathrm{S}}$,
 and
 (b): energy transfer for overall average
 and
 those for the active and moderate phases.
 \label{fig:phase}
 }
\end{figure}

The total-energy spectrum and the stretching-energy spectrum for the overall average
and those for the active and moderate phases
are drawn in Fig.~\ref{fig:phase}(a).
Obviously,
the intermittency results mainly from the fluctuation of the stretching energy
in the small and middle wave-number ranges ($k\lessapprox 400$),
i.e., in forcing and strong turbulence ranges.
No significant difference appears
in the large wave numbers, i.e., in the weak turbulence range,
which is consistent with the weak turbulence theory.
On the other hand,
the spectra of the kinetic and bending energies
are not so different between the active and moderate phases,
though they are not drawn here.

As examined in Ref.~\cite{PhysRevE.90.063004},
one-dimensionalized energy transfers,
$\mathcal{T}(k)=(\Delta k)^{-1}\sum_{|k^{\prime} - k|<\Delta k/2} T_{{\bm{k}^{\prime}}}$,
 at the active and moderate phases
are drawn
in Fig.~\ref{fig:phase}(b).
The energy transfer at the active phases 
shows negatively large values for the wave-number modes in the forcing range
and positive values for the wave-number modes in the strong turbulence range,
while the energy transfer at the moderate phases
is small negative in both of the wave-number ranges.
At these active phases,
the energy obtained from the external force in the forcing range
as the stretching energy
is distributed over the wave-number modes in the strong turbulence.
The energy transfer is relaxed to the normal state
at other phases.

\begin{figure}[t]
 \includegraphics[scale=.7]{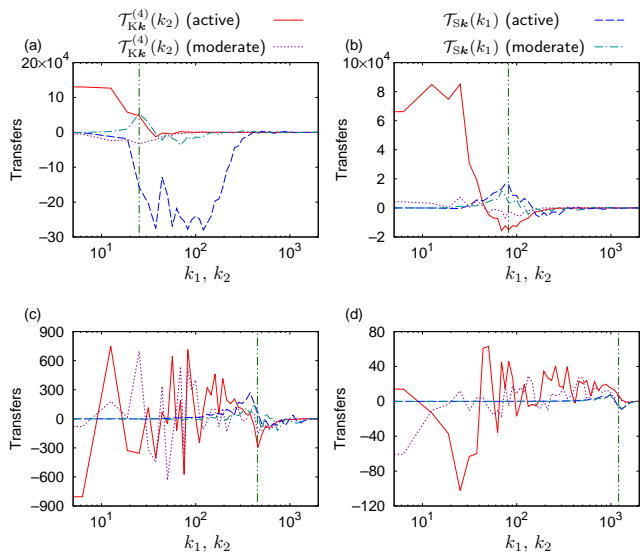}
 \caption{(Color online)
   Binary energy transfers at active and moderate phases.
   See the figure caption of Fig.~\ref{fig:intave}.
 \label{fig:intcond}
 }
\end{figure}
To identify the energy balance between two wave-number modes at these phases,
the binary energy transfers at the active phase and at the moderate phase
at the four representative wave numbers
are drawn in Fig.~\ref{fig:intcond}.
(See also Fig.~\ref{fig:intave}.)
The remarkable difference between these phases
is found 
at the wave number $|\bm{k}|=8\pi$ in the forcing range (Fig.~\ref{fig:intcond}(a))
and the wave number $|\bm{k}|=26\pi$ in the strong turbulence range (Fig.~\ref{fig:intcond}(b)).
At the wave number $|\bm{k}|=144\pi$ in the weak turbulence (Fig.~\ref{fig:intcond}(c)) only slight difference in finite range can be seen 
though it is hidden in the fluctuation:
$\mathcal{T}_{\mathrm{S} 144\pi}$ in the active phase is larger than that in the moderate phase
in $200 \lessapprox k_1 \lessapprox 400$,
and $\mathcal{T}_{\mathrm{K} 144\pi}$ in $100 \lessapprox k_2 \lessapprox 300$.
No clear difference within the fluctuation error can be observed
at the wave numbers $|\bm{k}|=384\pi$ in the dissipation range (Fig.~\ref{fig:intcond}(d)).

The nonlocal energy transfer from $V_{\mathrm{S} \bm{k}}$ where $\bm{k}$ is in the forcing range
($\mathcal{T}_{\mathrm{S}8\pi}(k_1)$ at the active phase in Fig.~\ref{fig:intcond}(a))
to $K_{\bm{k}}$ where $\bm{k}$ is in the strong turbulence 
($\mathcal{T}_{\mathrm{K}26\pi}(k_2)$ at the active phase in Fig.~\ref{fig:intcond}(b))
emerges only at the active phase.
It is clear that
the stretching energy in the forcing range
is distributed widely over the wave numbers in the strong turbulence
through the nonlocal energy transfer.
The local energy transfers between comparable-scale modes
are observed both at the active phases and the moderate phases
and at all the wave numbers.
The nonlocal energy transfer plays a crucial role only at the active phase
in the forcing range or in the strong turbulence.
In other words,
the local and nonlocal energy transfers coexist at the active phase,
while only the local energy transfer is effective at the moderate phase.

 \begin{figure}[t]
 \includegraphics[scale=.6]{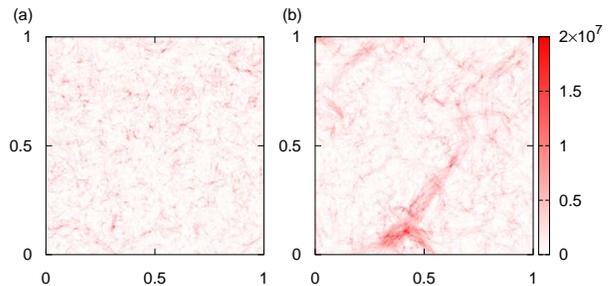}
 \caption{(Color online)
  Real-space structures in stretching energy fields at moderate phase (a)
  and at active phase (b)
  pointed out by arrows in Fig.~\ref{fig:tsnon}.
 \label{fig:stretchactmod}
 }
 \end{figure}

Next, energy transfer at each phase is examined
from the view point of real-space structures.
Since we have observed that
the stretching energy causes the intermittency,
the stretching energy in the real space,
which is defined as Eq.~(\ref{eq:VSreal}),
is drawn in Fig.~\ref{fig:stretchactmod}.
While the stretching-energy field at a moderate phase
pointed out by the left arrow of a pair of arrows in Fig.~\ref{fig:tsnon}
is drawn in Fig.~\ref{fig:stretchactmod}(a),
that at an active phase
pointed out by the right arrow is drawn in Fig.~\ref{fig:stretchactmod}(b)
as representative.
One can observe a distinctive structure
lying in the center of lower half part in this figure.
The distinctive structure is composed of two bundles of the fibrous structures,
and is in the form
of the laterally reversed image of the character $\lambda$.
A point-like structure appears at the joint of the two bundles,
i.e., roughly at $(0.4, 0.1)$.
The bundle structures and the point-like structures
appear intermittently,
and they are similarly observed in the field of the von Mises stress,
which is used for a criterion of yielding.
The slender elongated structures, i.e., the bundle structures,
are expected to reflect the nonlocal interaction of the energy transfer
at the active phases shown in Fig.~\ref{fig:intcond}(a),
since the lengths and the widths of such structures are, respectively,
$k\sim 2\pi\times O(1)$ and $k\sim 2\pi\times O(10)$.
The nonlocal property of energy transfer has been discussed also
in the Navier--Stokes turbulence,
where very thin and elongated intense vortices are responsible 
for anomalous corrections to the cascade theory.
In two-dimensional Navier--Stokes turbulence, especially,
reported are the nonlocality of the enstrophy cascade citing the coherent vortices
and the importance of the nonlocal feedback of small scales on large scales,
e.g., Ref.~\cite{doi:10.1063/1.857818,*PhysRevLett.83.4061}.

The point-like structures and the bundle structures might remind us of
the {\itshape d}-cones and the ridges of the displacement $\zeta$ in the real space
observed in laboratory experiments.
Although the point-like structures and the bundle structures are distinct
in the field of stretching energy,
the {\itshape d}-cones and the ridges could not be observed clearly in the field of $\zeta$ in the present simulation.
The point-like structures and the bundle structures
show the concentration of the stress,
and they would result in the {\itshape d}-cones and the ridges
in much higher nonlinear state,
because the relation between the structures in $\zeta$
and those in the energy dissipation field is reported 
in numerical simulations for strongly nonlinear field~\cite{PhysRevLett.111.054302}.

It should be noted that
though there are no large-scale bundle structures,
small unilaminate fibrous structures and resulting point structures
can be observed
even at the moderate phase (Fig.~\ref{fig:stretchactmod}(a)),
which might be related to recently-reported small-scale intermittency~\cite{PhysRevE.94.011101}.
At both phases,
small-scale fibrous structures are found also in the gradient field of displacement $|\nabla \zeta|$~\cite{PhysRevLett.111.054302},
though the magnitude at the active phase is much larger than that at the moderate phase.
These strongly-nonlinear structures that dynamically emerge
are observed only in numerical experiments.
The observation in laboratory experiments is expected.

\begin{figure}[t]
 \includegraphics[scale=.75]{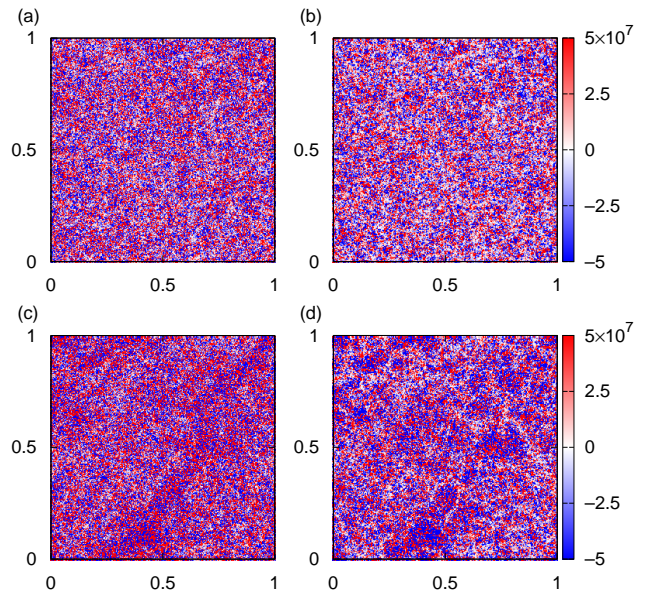}
 \caption{
 (Color)
 Energy transfers in real space.
 (a) and (b): at moderate phase,
 and 
 (c) and (d): at active phase.
 (a) and (c): to kinetic energy $T_{\mathrm{K}}^{(4)}(\bm{x})$,
 and 
 (b) and (d): to stretching energy $T_{\mathrm{S}}(\bm{x})$.
 \label{fig:transferinreal}
 }
\end{figure}

The energy transfers at each location in the real space
defined as Eqs.~(\ref{eq:transferinRK}) and (\ref{eq:transferinRS})
are drawn in Fig.~\ref{fig:transferinreal}.
As a representative of the moderate phase, 
Figs.~\ref{fig:transferinreal}(a) and (b) are drawn
at the same time of Fig.~\ref{fig:stretchactmod}(a).
Similarly,
Figs.~\ref{fig:transferinreal}(c) and (d)
correspond to Fig.~\ref{fig:stretchactmod}(b) in time.
At the active phase,
the energy transfers are large around the bundle of the fibrous structures
near $(0.5, 0.2)$,
though it might be difficult to be recognized owing to the numerous small-scale variation.

\begin{figure}[t]
 \includegraphics[scale=.75]{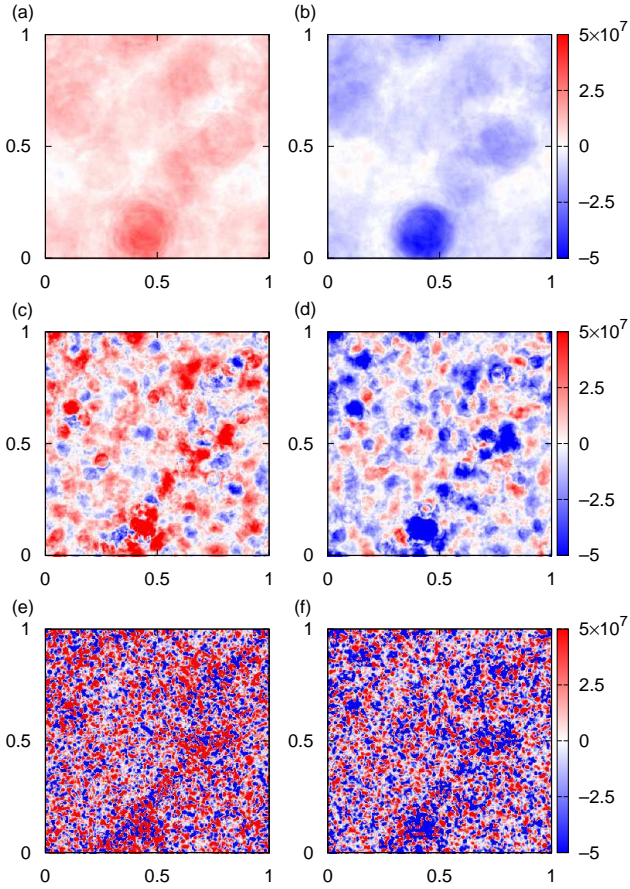}
 \caption{
 (Color)
 Coarse-grained energy transfers in real space at active phase.
 (a) and (b): $r_{\mathrm{CG}} = 1/4$, i.e., $k_{\mathrm{CG}}=8\pi$.
 (c) and (d): $r_{\mathrm{CG}} = 1/16$, i.e., $k_{\mathrm{CG}}=32\pi$.
 (e) and (f): $r_{\mathrm{CG}} = 1/64$, i.e., $k_{\mathrm{CG}}=128\pi$.
 (a), (c) and (e): to kinetic energy $T_{\mathcal{K}}^{(4)}(\bm{x})$,
 and 
 (b), (d) and (f): to stretching energy $T_{\mathcal{S}}(\bm{x})$.
 \label{fig:coarsetransferinrealactive}
 }
\end{figure}

To observe the energy transfer in the real space more clearly,
the coarse-graining of the finite area $A$ in \S\ref{ssec:formulationR}
is employed for the fields in Fig.~\ref{fig:coarsetransferinrealactive}.
The energy transfers
which are coarse-grained with the scale $r_{\mathrm{CG}}$
 at the active phase are drawn
in Fig.~\ref{fig:coarsetransferinrealactive}.
In Figs.~\ref{fig:coarsetransferinrealactive}(a) and (b),
the coarse-grained transfers are averaged over the grid points $\bm{x}^{\prime}$
in $|\bm{x}^{\prime} - \bm{x}| < r_{\mathrm{CG}} = 1/4$.
The scale corresponds to $k_{\mathrm{CG}}=8\pi$ in the forcing range.
Similarly,
the coarse-grained transfers with the scale $r_{\mathrm{CG}} = 1/16$,
and those with the scale $r_{\mathrm{CG}} = 1/64$
are drawn 
in Figs.~\ref{fig:coarsetransferinrealactive}(c) and (d),
and
in Figs.~\ref{fig:coarsetransferinrealactive}(e) and (f),
respectively.
The scale $r_{\mathrm{CG}} = 1/16$ and the scale $r_{\mathrm{CG}} = 1/64$,
respectively,
correspond
to the wave number $k_{\mathrm{CG}}=32\pi$ in the strong turbulence
and to the wave number $k_{\mathrm{CG}}=128\pi$ in the weak turbulence.

In the coarse-grained transfers with the scale $r_{\mathrm{CG}}=1/4$ (Figs.~\ref{fig:coarsetransferinrealactive}(a) and (b)),
it is found that
the bundle structures play a dominant role in the energy transfer;
the scale gains the kinetic energy,
and loses the stretching energy.
The coarse-grained transfer of the kinetic energy with the scale $r_{\mathrm{CG}}=1/4$
is positive almost everywhere,
while that of the stretching energy is negative.
They are consistent with the energy transfers in the Fourier space
as shown in Fig.~\ref{fig:intave}(a).

In the coarse-grained transfers with the scale $r_{\mathrm{CG}}=1/16$ (Figs.~\ref{fig:coarsetransferinrealactive}(c) and (d)),
the energy transfers in the vicinity of the bundle structures
are a little less remarkable,
and the positive and negative regions are mixed in confusion
 in the region away from the bundle structures.
However, the kinetic-energy and stretching-energy transfers,
respectively, tend to positive and negative near the bundle structure
in this field.
In fact, the kinetic-energy and stretching-energy transfers
are positive and negative on average, respectively.

The coarse-grained transfers with the scale $r_{\mathrm{CG}}=1/64$ (Figs.~\ref{fig:coarsetransferinrealactive}(e) and (f))
are similar with those without coarse-graining
as shown in Fig.~\ref{fig:transferinreal}.
The large energy transfers in the bundle structures
can be recognized with an effort.

At the moderate phase,
the regions of the positive and negative transfers
are mixed for any scales,
though the figure is omitted here.
Thus, no characteristic structures are found 
as anticipated from Fig.~\ref{fig:stretchactmod}(a).
Even in the largest scale $r_{\mathrm{CG}}=1/4$,
both the negative and the positive regions of energy transfers appear
in contrast with those at the active phase shown in Figs.~\ref{fig:coarsetransferinrealactive}(a) and (b).
Although no distinctive structures can be observed in all the contours of Fig.~\ref{fig:coarsetransferinrealactive},
The negagive correlations between $T_{\mathcal{K}}^{(4)}(\bm{x})$ and $T_{\mathcal{S}}(\bm{x})$
are large for all scales
as recognized from color distribution.
It suggests that the energy transfer through boundaries of $A$ is small as expected.

\begin{figure}[t]
 \includegraphics[scale=.7]{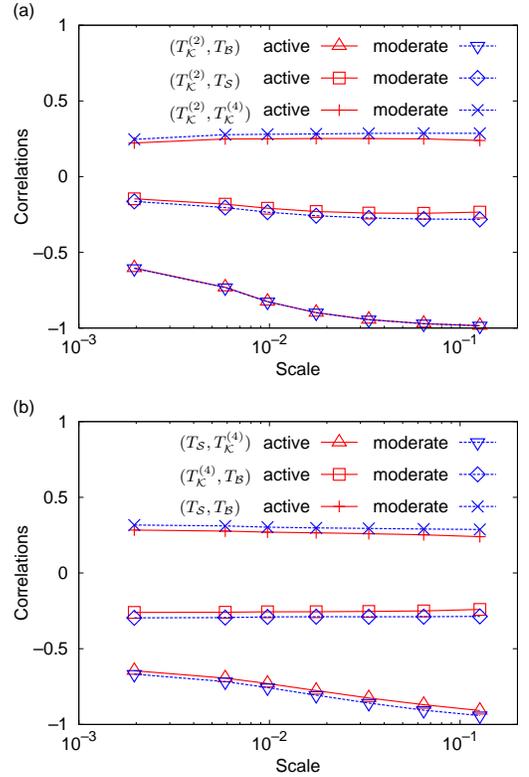}
 \caption{(Color online)
 Scale dependence of the correlations between energy transfers.
 \label{fig:coarsegraining}
 } 
\end{figure}

To quantify the scale-dependence of the above correlation,
the correlations between energy transfers defined by
\begin{align}
   C(r_{\mathrm{CG}}; T_X, T_Y)
 = \frac{\langle (T_{\mathcal{X}}-\langle T_{\mathcal{X}} \rangle)
                 (T_{\mathcal{Y}}-\langle T_{\mathcal{Y}} \rangle) \rangle}%
 {\sqrt{\langle (T_{\mathcal{X}}-\langle T_{\mathcal{X}} \rangle)^2 \rangle 
        \langle (T_{\mathcal{Y}}-\langle T_{\mathcal{Y}} \rangle)^2 \rangle}}
\end{align}
are shown in Fig.~\ref{fig:coarsegraining},
where $T_{\mathcal{X}}$ and $T_{\mathcal{Y}}$ are coarse-grained field of 
$T_X$ and $T_Y$ with the scale $r_{\mathrm{CG}}$, respectively. 
Here, $\langle\cdot\rangle$ represents averaging over space and ensembles.

Only two combinations of energy transfers, 
namely $(T_{\mathcal{K}}^{(2)}, T_{\mathcal{B}})$ and $(T_{\mathcal{S}}, T_{\mathcal{K}}^{(4)})$,
show large correlation and its scale-dependence clearly.
These correlations are negative 
and decrease toward $-1$ as the coarse-grained scale become large.
The absolute values of correlations at the moderate phases are larger
than those at the active phases,
which might be understood by recalling the fact that
the interactions among different scales are larger at the active phases.

It might be interesting to point out that
the correlation between linear energy transfers $(T_{\mathcal{K}}^{(2)}, T_{\mathcal{B}})$
approaches $-1$ with increasing scales 
faster than the one between nonlinear energy transfers $(T_{\mathcal{S}}, T_{\mathcal{K}}^{(4)})$.
The coexistence of the weak turbulence and the strong turbulence 
has been reported in the previous papers~\cite{PhysRevLett.110.105501,PhysRevE.89.012909,PhysRevE.90.063004};
the weak turbulence in the small scales, i.e., the large wave numbers,
and the strong turbulence in the large scales, i.e., the small wave numbers.
In other words, 
the energy transmutation between the kinetic energy and the bending energy
via $(T_{\mathcal{K}}^{(2)}, T_{\mathcal{B}})$ is active in both weak and strong turbulence,
while the energy transfer between the kinetic energy and the stretching energy
via $(T_{\mathcal{S}}, T_{\mathcal{K}}^{(4)})$ is mainly in the strong turbulence.

 \section{Summary}

We have numerically analyzed the energy transfers
in the statistically-steady non-equilibrium state
in elastic-wave turbulence.
We have focused on the nonlinear kinetic-energy transfer $T_{\mathrm{K}}^{(4)}$ and the nonlinear stretching-energy transfer $T_{\mathrm{S}}$,
since
the energy transfers among different scales occur only through 
the exchange between the kinetic energy and the stretching energy
as reported in Ref.~\cite{PhysRevE.90.063004}.
In other words,
the analytical expression of the energy transfers declares that
the wave-number modes of the bending energy are only a mediator in the triad interaction.
We have successfully visualized
the energy transfer through a triad interaction,
 and hence the transfer between the kinetic energy of a wave-number mode
 and the stretching energy of another wave-number mode
by using the analytical expression of the detailed energy balance among three wave-number modes.
The local and nonlocal interactions of energy transfers can be recognized.
The nonlocality of the net energy transfers in the strong turbulence has been found.
The energy transfers from or to the representative scales
were shown,
and they reveal the distinctive configuration of the interactions
in each scale of the forcing, strong-turbulence, weak-turbulence, and dissipation ranges.

By using the time variation of the systems's nonlinearity,
the energy transfers were investigated separately in the active and moderate phases.
At the active phases,
the bundles of fibrous structures and the point-like structures
emerge in the real space,
and these intermittent real-space structures cause a large amount of the energy transfer in the Fourier space then.
At the moderate phase,
no such strongly nonlinear structures have been observed,
and the energy transfers are caused mainly by the weakly nonlinear interactions
of elementary waves as described in the weak turbulence theory.

\begin{figure}[t]
\includegraphics[scale=.65]{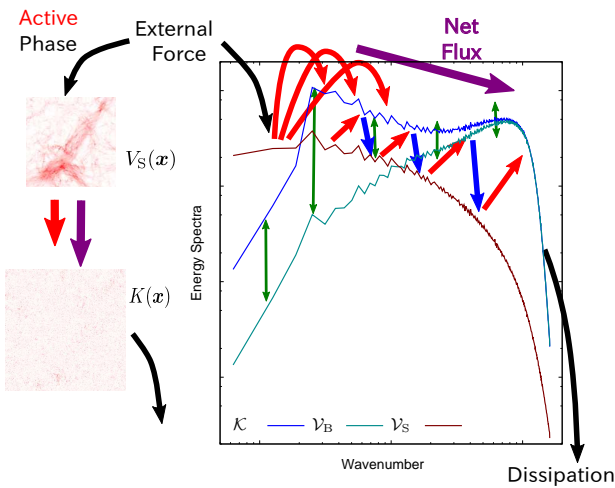}
 \caption{(Color online)
 Schematic representation of energy transfers.
 \label{fig:schematicenergytransfer}
 }
\end{figure}

In Fig.~\ref{fig:schematicenergytransfer},
we have schematically obtained an integrated picture of the energy transfers
in the elastic-wave turbulence maintained by large-scale forcing.
The external force gives energy to the forced wave numbers
as its stretching energy at the active phase.
In the real space,
the strongly-nonlinear intermittent bundle structures are created at the active phase.
The intermittent structures distribute the energy to the wave numbers
in the strong turbulence
through the nonlocal interactions in the Fourier space.
The locally-cascading interactions between the kinetic and stretching energies
also transfers energy to the large wave numbers.
In the weak turbulence,
the energy transfer due to the nonlocal interactions
vanishes,
and the weakly-nonlinear resonant interactions according to the weak turbulence theory,
which are local in the Fourier space,
play a dominant role.
After the energy is transferred to the dissipation range,
the energy is dissipated from the kinetic energy.
The nonlocal interactions due to the intermittent structures
and the locally-cascading interactions between the kinetic and stretching energies
make the net energy flux to the large wave numbers.
The bending energy mediates the energy transfer in the nonlinear interactions,
and it is directly involved only in the linear transmutation between the kinetic energy and the bending energy.

\begin{acknowledgments}
 Numerical computation in this work was carried out
 using the computer facilities
 at the Yukawa Institute, Kyoto University
 and Research Institute for Information Technology, Kyushu University.
 This work was partially supported by KAKENHI Grant
 No.~15K17971, No.~16K05490 and No.~17H02860.
\end{acknowledgments}

\end{document}